\documentclass[12pt,preprint]{aastex}

\newcommand{\kms}{\ km\,s$^{-1}$}
\newcommand{\lsun}{\mbox{\ $L_{\odot}$}}

\slugcomment{Submitted to the Astrophysical Journal Letters}

\shorttitle{Water Maser at $z=0.66}
\shortauthors{Barvainis}

\begin{document}

\title{Extremely Luminous Water Vapor Emission from a Type 2 Quasar at Redshift $z = 0.66$} 

\author{Richard Barvainis\footnote{Any opinions, findings, conclusions, and recommendations expressed
in this material are those of the author and do not necessarily reflect the views of the National 
Science Foundation.}}
\affil{National Science Foundation, Arlington, VA 22230 }
\affil{Physics Department, Gettysburg College, Gettysburg, PA 17325}
\affil{Email: rbarvai@nsf.gov}

\author{Robert Antonucci}
\affil{Department of Physics, UC Santa Barbara, Santa Barbara, CA 93106}
\affil{Email: ski@physics.ucsb.edu}
\begin{abstract}
A search for water masers in 47 Sloan Digital Sky Survey Type 2 quasars using the
Green Bank Telescope has yielded a detection at a redshift of $z = 0.660$.  This maser is more
than an order of magnitude higher in redshift than any previously known and,
with a total isotropic luminosity of  23,000\lsun, also the most
powerful.  The presence and detectability of water masers in
quasars at $z\sim 0.3-0.8$ may provide a better understanding of quasar molecular 
tori and disks, as well as fundamental quasar and galaxy properties such as black hole masses.  
Water masers at cosmologically interesting distances may also eventually provide, via direct 
distance determinations, a new cosmological 
observable for testing the reality and properties of dark energy, currently inferred 
primarily through Type 1a supernova measurements.

\end{abstract}


\section{Introduction}

Water vapor emission in the form of powerful masers has been detected
in $\sim 10$\% of Type 2 Active Galactic Nuclie (AGNs) observed in the local universe
\citep{braatz04}.  
Some of these masers have spectral characteristics suggesting that they 
arise in a dusty molecular 
disk or torus viewed edge-on, where conditions are favorable for their formation.
Such masers have provided unique diagnostics of some of the basic characteristics of AGNs, 
including information on the central black hole masses.  Furthermore, in one case, 
the Seyfert 2 / LINER galaxy 
NGC4258, water masers have been used to 
determine a direct geometrical distance \citep{herrn99} to the galaxy independent of the assumptions 
of the standard extragalactic distance ladder \citep{madore99}. 

Most previous searches for water masers in AGNs have been confined to relatively 
nearby objects such as Seyfert galaxies, LINERs, and radio galaxies at 
$z << 0.1$ \citep{braatz04,tarchi03}.  
The vast majority of water masers have been found in Type 2 AGNs and LINERS.  These are systems where
the disk or torus thought to surround the nucleus in most AGNs is seen edge on, obscuring
the nucleus optically but allowing masers beamed in the midplane of the torus
to be observed at Earth.  Until recently, the only Type 2 objects known at higher redshifts
($z > 0.1$) were radio galaxies and a few Ultraluminous Infrared Galaxies; searches
for maser emission in high-$z$ radio galaxies to date have been unsuccessful (Barvainis, Henkel, 
\& Antonucci, unpublished).

The recent compilation of a large new sample of luminous 
Type 2 quasars at redshifts between 0.30 and 
0.83 by \citet{zakamska03} using the Sloan digital Sky Survey (SDSS) has provided a unique 
new opportunity to search for water maser emission in the distant universe.  This redshift
range corresponds closely to that of most of the Type 1a supernovae used to measure 
luminosity distance as a function of redshift, thereby testing cosmological models.  A 
result of fundamental importance to modern physics is the determination, based
on the supernova studies, that the content of the universe may be dominated by dark 
energy, a poorly understood quantity that may be related to Einstein's Cosmological Constant. 
However, the possibility remains that the Type 1a supernova measurements may be subject to 
systematic errors \citep{riess04}, and an independent angular size distance measurement on 
cosmological scales to check the supernova results would be highly desirable. 

Here we report the discovery of 
a water maser at redshift $z = 0.660$ in the Type 2 Quasar SDSS  J080430.99+360718.1 (hereafter
J0804+3607) using the Green Bank Telescope (GBT).  This is the most distant water maser 
known by over an order of magnitude, 
and, with an apparent (isotropic) luminosity of $2.3\times10^4$\lsun, by far the most powerful.
In addition to its implications for the study of black hole masses and other 
properties of quasars, this detection shows that powerful masers exist at redshifts 
high enough to potentially test the existence of dark energy as derived from studies of Type 1a 
supernovae.

\section{Observations}

The observations were carried out on January 6 and January 9
2005 using the Green Bank Telescope of the National Radio Astronomy Observatory.  
The line was detected in both sessions (see Figure 1 for the final sum of 
both nights).  No line was detected in the spectrum of another quasar at similar redshift, so the 
line cannot be  attributed to interference or a receiver ``birdie".  A total of 47 SDSS Type 2 
quasars were  observed, with J0804+3607 yielding the only detection. The sample is from 
\citet{zakamska03}. 

Each quasar was observed for 28 minutes of integration time in the initial survey. We obtained 
an additional 164 minutes (after discarding one bad 4-minute scan) of follow-up time on J0804+3607, 
for a total  integration time of 
192 minutes.  The observing mode utilized two feeds separated by 5.5 arcminutes on the sky, 
each with dual  polarization.  The source was placed alternately in each beam, with a 
position switching interval of 2 minutes.  Antenna pointing checks were made roughly every 2 hours, 
and typical pointing errors were less than 1/10 of a beamwidth.  The antenna beamwidth is 
1 arcminute at 13.4 GHz, the frequency of the detected line.   We estimate the calibration 
uncertainty to be $\sim 20$\%.                                 

The spectrum shown in Figure 1 has 
been hanning smoothed once to provide a channel width of 0.1 Mhz, 
or a velocity channel width of 2.2\kms.  The RMS noise on the baseline is 0.48 mJy.
The central frequency of the main spike is 13.391 GHz, 
yielding a redshift of 0.660.  
The original spectrum was 200 MHz wide, and  
evidenced no other line features outside the region shown in the figure.   

We discount a false identification with a local source of the J=7/2, F=4-4 13.44 GHz OH 
maser (e.g., Baudry \& Diamond 1998) for the following reasons: a) such masers are rare;
b) they are associated with regions of star-formation and as such will normally be found 
only towards the galactic plane;  c) the probability of the detection of a slightly red-shifted 
13.44 GHz OH maser within the beam of the GBT in a random direction is very low; and d) assuming
the line to be water the redshift is extremely close, within 0.002, to that of the quasar  
($z = 0.658$ based on narrow optical emission lines).

\section{Discussion}

Although the maser characteristics of J0804+3607 may or may not lend themselves to a direct 
distance determination as in the case of the nearby NGC4258, our discovery of a maser in the 
range $0.5 < z < 1$ opens up the possibility of finding objects like NGC4258 that 
could be used to measure cosmological distances and test dark energy.  
The maser spectrum of J0804+3607 (Figure 1) consists of a narrow spike 
with a full width at half maximum (FWHM) of about 10\kms\  
and a full width at zero intensity of about 40\kms, together with 
a possible weak 
underlying broader component or wing extending to lower frequencies with FWHM $\sim 100$\kms.  

NGC4258 has a complex system of variable features spanning about 100\kms\ at the 
systemic velocity, with satellite lines centered at $\pm 900$\kms\ having maximum flux 
densities of $15-40$\% (time variable) of the main peak.   
In J0804+3607 there is no significant evidence for satellite features separated from 
the main emission. Any such lines would have been detectable at 40\% of the main peak but not at 
15\%, given the signal-to-noise ratio obtained.  

The redshift of the maser line in J0804+3607 is $z = 0.660$, slightly higher than the optical 
redshift of $z = 0.658$ based on O[III] emission.  The line is thus redshifted by 
360\kms\ relative to the O[III] velocity; there is no way to tell whether the detected
maser component is a ``systemic'' or ``satellite" (i.e., high-velocity) feature in a disk geometry.   
The maser is more than an order of magnitude higher in redshift than
any previously known water maser -- 3C403 was the highest at $z = 0.059$ \citep{tarchi03}.  
The total luminosity of the maser emission is 23,100\lsun,
assuming $H_0 = 71$\kms Mpc$^{-1}$, $\Omega_M = 0.27$, and $\Omega_{\Lambda} = 0.73$. 
This includes both the narrow spike and the weak broader
component.  The next highest known maser luminosity is 6,800\lsun\  
for TXS 2226-184 \citep{koek95} (calculated using the same cosmological parameters).

The use of extragalactic H$_2$O masers as physical probes of the inner regions
of AGNs saw a spectacular payoff in NGC4258 \citep{miyoshi95,greenhill95},
where VLBI followup of
the initial  single-dish discovery revealed precise Keplerian rotation about a massive ($10^7
M_{\sun}$) compact object.  
Very importantly for AGN studies, the black hole mass was
determined directly for the first time.  Similarly, the masers in NGC1068 are in 
nearly Keplerian rotation and an upper limit to the black hole mass derived from the 
masers led to the fundamental conclusion that the continuum luminosity of the AGN is 
of order the Eddington luminosity \citep{greenhill96}. 

Simple and accurate dynamical arguments for NGC4258 produced a distance
of $7.2\pm 0.3$ Mpc \citep{herrn99}.  Although very useful for calibrating the distance ladder, 
this is too nearby to be indicative of the Hubble constant.  A 
direct measurement of $H_0$ would be possible if a maser system like NGC4258
were found at a distance that places it well into the Hubble flow, and
some candidates for this already exist \citep{braatz04}.
One goal of the present study is to take this further and see what can be found at 
redshifts large enough that the apparent universal acceleration/cosmological constant/dark 
energy could be explored. 
If maser distances can be determined at high redshifts ($z > 0.4$),
deviations from the Hubble law could be detected similar to
those claimed by the supernova cosmology teams \citep{perlmutter99,garnavich98,riess04}.   
The assertions by these teams of universal acceleration based on Type Ia supernovae, and concomitant
evidence for dark
energy, clearly need observational confirmation and elaboration.  The technique outlined
here could, with some luck, provide such a check.  Even a measurement of the angular
size distance to a single object to $\sim 10\%$ precision could help to distinguish between  
different cosmologies.  For example, 
the ratio of the angular size distance versus redshift curves at $z = 0.66$ for the Concordance
Cosmology ($\Omega_M = 0.27$, and $\Omega_{\Lambda} = 0.73$) and a matter-dominated
flat cosmology ($\Omega_M = 1$) is ${1437\ {\rm Mpc} / 1139\ {\rm Mpc}} = 1.26$.

So far we have observed 47 objects, but there are 
at least 270 other SDSS Type 2 quasars known that are accessable via the 
frequency coverage of large telescopes
like the GBT and the 100-m Effelsberg antenna.  We are pursuing a much larger 
survey in search of appropriate maser systems for direct cosmological distance 
determinations.
 
The H$_2$O megamaser theory of \citet{neuf94}
predicts that roughly 100\lsun\  of maser
emission is produced for each square parsec of area illuminated by the primary
AGN x-ray emission.  If the maser emission is anisotropic, e.g.\ beamed in the 
plane of the torus, the {\it apparent} luminosity could be much larger. 
Studies of the infrared luminosity from dust in quasars (e.g., Haas et al.\ 2000) 
suggest that the area of warm, dusty, illuminated molecular clouds increases in
rough proportion to optical/UV/x-ray luminosity:  the sublimation
radius of the dust, which sets the innermost possible radius for the torus,
scales as $L_{\rm opt/UV}^{1/2}$, where $L_{\rm opt/UV}$ is the optical/UV
continuum luminosity \citep{barv87}, and thus the {\it area} of the torus
illuminated scales simply as $L_{\rm opt/UV}$.  Extremely powerful masers
might thus be expected from high-luminosity quasars. Our discovery supports 
this scenario.  Later theory papers made the geometry assumed by 
\citet{neuf94} more realistic, 
and considered various energy sources for the maser.  These models, which do not 
change the basic energetics, are reviewed in \citet{maloney02}. 

If we assume that the masers in J0804+3607 are analogous to those in NGC4258, we can 
compute their approximate distance from the central optical/UV/x-ray source of the quasar.  
Making a rough estimate of the continuum luminosity based on the O[III] $\lambda 5007$
line flux using typical Type 1 quasar ratios \citep{zakamska03,elvis94}, we find a 
bolometric luminosity of $L_{\rm bol} \sim 2\times 10^{45}$ erg~s$^{-1}$ for J0804+3607.  
For NGC4258, $L_{\rm bol} \sim 10^{42}$ erg~s$^{-1}$ \citep{kart99}.  The radiative flux 
impinging on the masing region, which controls the physical conditions in the gas 
(especially the x-ray flux), scales as $L_{\rm bol} / r^2$.  
With this and the relative bolometric luminosities the radius of the maser region would 
be 45 times larger in
J0804+3607 than in NGC4258.  Given that the masers in NGC4258 arise over a range 
$0.14 < r < 0.28$ pc \citep{herrn05}, the range in J0804+3607 would be  
$6.3 < r < 12.6$ pc.  For the 
adopted cosmology this translates to an angular diameter of the masing region of $2-4$ milliarcsec,
which is easily resolvable using VLBI.   

  In summary, this result indicates that water masers are detectable at high redshift, 
potentially providing information on black hole masses and physical conditions in the 
masing regions (molecular torus / warped disk) of quasars.  High redshift masers might 
also eventually provide angular 
size distances as a function of redshift, a new cosmological observable.  

\acknowledgements{The National Radio Astronomy Observatory is a facility of the National
Science Foundation operated under cooperative agreement by Associated 
Universities, Inc.  The authors wish to thank Yuri Kovalev for observing help at the GBT, 
David Neufeld for advice on maser theories, and Jim Braatz 
for valuable comments on the 
manuscript. The referee alerted us to the existence of the the 13.44 GHz OH line and provided
the reasons why it was an unlikely identification for the observed line, for which we are grateful.}

\begin{figure*}
\figurenum{1}
\epsscale{1.2}
\includegraphics[height=16cm, angle=-90]{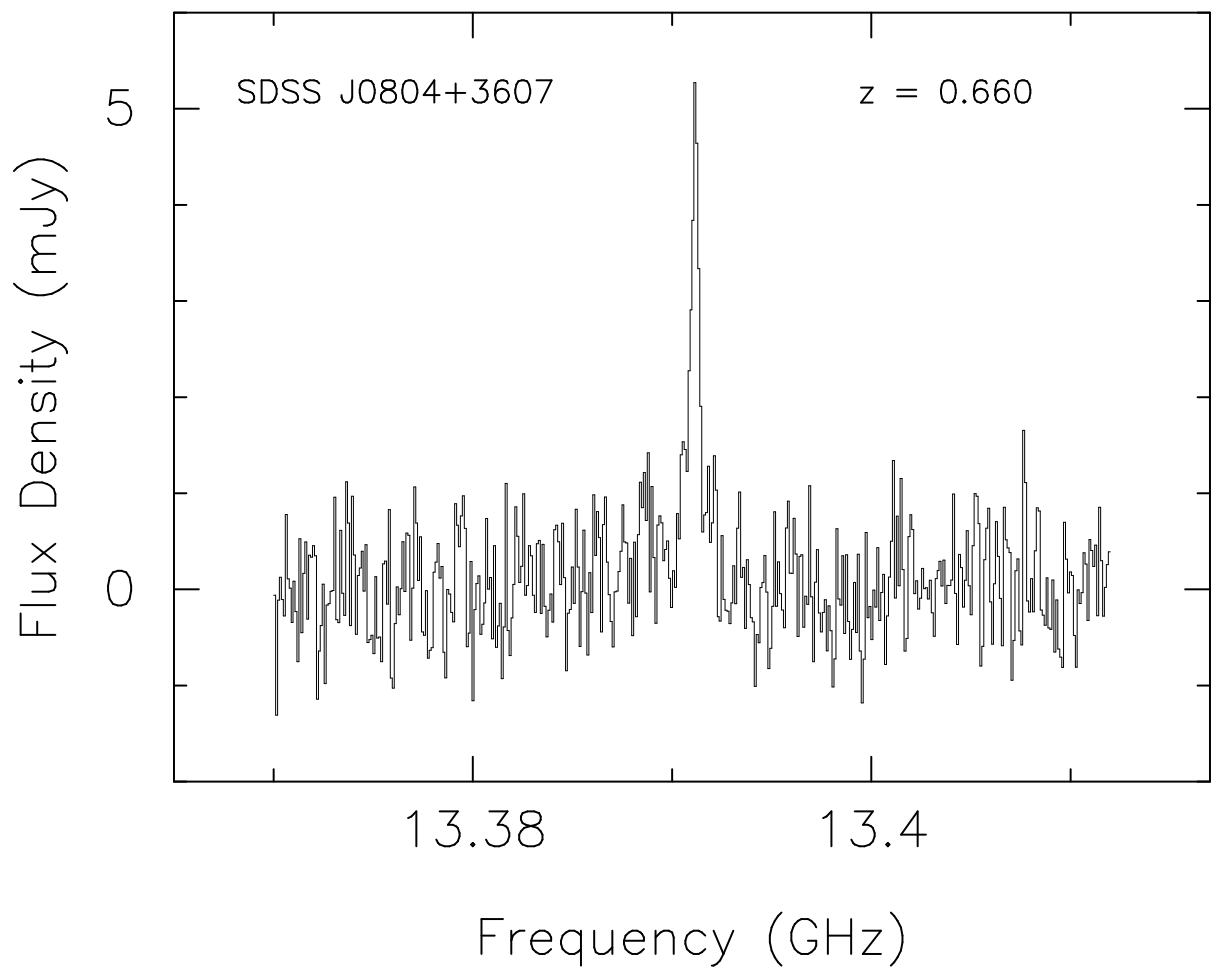}
\caption {Spectrum of water maser emission at $z=0.660$ from the Type 2
quasar SDSS J080430.99+360718.1.  The full bandwidth of 200 MHz is 
not shown here to better emphasize the line.   The effective spectral resolution 
and channel width, after hanning smoothing, is 0.1 MHz, or 2.2 km s$^{-1}$.}
\end{figure*}


\begin{thebibliography}{}
\bibitem [Barvainis(1987)]{barv87} 
Barvainis, R.  1987, ApJ, 320, 537
\bibitem [Baudry \& Diamond(1998)]{baudry98}
Baudry, A., \& Diamond, P.J. 1998, A\&A, 331, 697
\bibitem [Braatz et al.(2004)]{braatz04} 
Braatz, J. A., Henkel, C., Greenhill, L. J., Moran, J. M., \& Wilson, A. S. 
2004, ApJ, 617, L29
\bibitem [Elvis et al.(1994)]{elvis94}
Elvis, M.\ et al.  1994, ApJSupp, 95, 1
\bibitem [Garnavich et al.(1998)]{garnavich98} 
Garnavich, P.M., et al.  1998, ApJ, 509, 74
\bibitem [Greenhill et al.(1995)]{greenhill95} 
Greenhill, L.~G., Jiang, R.~D., Moran, J.~M., Reid, M.~J., Lo, K.~Y., \& Claussen, M. J.  1995, ApJ, 440, 619
\bibitem [Greenhill et al.(1996)]{greenhill96} 
Greenhill, L. J., Gwinn, C. R., Antonucci, R., \& Barvainis, R. 1996, ApJ, 472, L21
\bibitem [Haas et al.(2000)]{haas00}
Haas, M. 2000, A\&A, 354, 453
\bibitem [Herrnstein et al.(1999)]{herrn99}
Herrnstein, J. R., Moran, J. M., Greenhill, L. J., Diamond, P. J., Inoue, M.,
Nakai, N., Miyoshi, M., Henkel, C., \&  Riess, A. 1999, Nature, 500, 539
\bibitem [Herrnstein et al.(2005)]{herrn05}
Herrnstein, J. R., Moran, J. M., Greenhill, \& Trotter, A.S. 2005, Astro-ph/0504405
\bibitem [Kartje, K\"onigl, \& Elitzur(1999)]{kart99}
Kartje, A.F., K\"onigl, A., \& Elitzur, M. 1999, 513, 180 
\bibitem [Koekemoer et al.(1995)]{koek95} 
Koekemoer, A. M., Henkel, C., Greenhill, L.J., Dey, A., van Breugel, W., Codella, C., \&  
Antonucci, R. 1995, Nature, 378, 697
\bibitem [Madore et al.(1999)]{madore99}
  Madore, B.~F., et al. 1999, ApJ, 515, 29 
\bibitem [Maloney(2002)]{maloney02}
Maloney, P.  2002, PASA, 19, 401
\bibitem [Myoshi et al.(1995)]{miyoshi95} 
Miyoshi, M., et al.  1995, Nature, 373, 127
\bibitem [Neufeld, Maloney, \& Conger(1994)]{neuf94}
Neufeld, D., Maloney, P., \& Conger, S.  1994, ApJ, 436, L127
\bibitem [Perlmutter et al.(1999)]{perlmutter99} 
Perlmutter, S., et al.  1999, ApJ, 517, 565
\bibitem [Riess et al.(2004)]{riess04} 
Riess, A.G., et al. 2004, ApJ, 607, 665
\bibitem [Tarchi et al.(2003)]{tarchi03} 
Tarchi, A., Henkel, C., Chiaberge, M., \& Menten, K. M. 2003, ApJ, 407, L33
\bibitem [Zakamska et al.(2003)]{zakamska03} 
Zakamska, N.L., et al. 2003, AJ, 126, 2125
\end{thebibliography}
\end{document}